\documentstyle[12pt]{article}
\newcommand{\ncm}{\newcommand}
\ncm{\oH}{\bar{H}}
\ncm{\us}{\quad\mbox{using}\quad}
\ncm{\ra}{\rightarrow}
\ncm{\ot}{\otimes}
\ncm{\DH}{D(H)}
\ncm{\DW}{D^{\omega}(H)}
\ncm{\TH}{T(\oH)}
\ncm{{\im}}{\imath}
\ncm{\ba}{\begin{array}}
\ncm{\ea}{\end{array}}
\ncm{\ul}{\underline}
\ncm{\ol}{\overline}

\def\hoek{\hbox{\vrule height 2.5ex depth 0pt \vrule width 2.5ex height .4pt
 depth 0pt}}
\def\haak#1#2{
\mathop{\hoek\llap{\vbox to 2.5ex{ \vfil
\hbox{$\scriptstyle#1$\hskip 2.8ex} \vfil}}}
\limits_{#2} }
\def\hook#1#2{\setbox0=\hbox{$\scriptstyle#1$}
\hskip\wd0\haak{\box0}{#2}}
\ncm{\str}{\rule{0cm}{3.5mm}}
\ncm{\om}{\omega}
\ncm{\ep}{\epsilon}
\newlength{\extraspace}
\newlength{\extraspaces}
\newcommand{\be}{\begin{equation}
\addtolength{\abovedisplayskip}{\extraspaces}
\addtolength{\belowdisplayskip}{\extraspaces}
\addtolength{\abovedisplayshortskip}{\extraspace}
\addtolength{\belowdisplayshortskip}{\extraspace}}
\newcommand{\ee}{\end{equation}}

\newcommand{\bea}{\begin{eqnarray}
\addtolength{\abovedisplayskip}{\extraspaces}
\addtolength{\belowdisplayskip}{\extraspaces}
\addtolength{\abovedisplayshortskip}{\extraspace}
\addtolength{\belowdisplayshortskip}{\extraspace}}
\newcommand{\eea}{\end{eqnarray}}
\newcommand{\beas}{\begin{eqnarray*}
\addtolength{\abovedisplayskip}{\extraspaces}
\addtolength{\belowdisplayskip}{\extraspaces}
\addtolength{\abovedisplayshortskip}{\extraspace}
\addtolength{\belowdisplayshortskip}{\extraspace}}
\newcommand{\eeas}{\end{eqnarray*}}

\ncm{\al}{\alpha}
\ncm{\bt}{\beta}
\ncm{\gm}{\gamma}
\ncm{\dl}{\delta}
\ncm{\varep}{\varepsilon}
\ncm{\zt}{\zeta}
\ncm{\et}{\eta}
\ncm{\th}{\theta}
\ncm{\kp}{\kappa}
\ncm{\lm}{\lambda}
\ncm{\rh}{\rho}
\ncm{\hl}{\hline}
\ncm{\sg}{\sigma}
\ncm{\ta}{\tau}
\ncm{\ph}{\phi}
\ncm{\phv}{\varphi}
\ncm{\ch}{\chi}
\ncm{\ps}{\psi}
\ncm{\nn}{\nonumber}
\title{Quantumgroups in the Higgs phase\thanks{Invited talk
presented by Mark de Wild Propitius
at `The III International Conference on Mathematical Physics,
String Theory and Quantum Gravity',
Alushta, Ukraine, June 13-24, 1993.}}
\author{ \small
\\F.Alexander Bais\thanks{bais@phys.uva.nl} $\:$  and
Mark de Wild Propitius\thanks{mdwp@phys.uva.nl}
\\ Instituut voor Theoretische Fysica\\Valckenierstraat 65,
1018XE Amsterdam, The Netherlands}
\date{November 1993}
\begin{document}
\maketitle
\begin{abstract}
In the Higgs phase we may be left with a residual finite symmetry group $H$
of the condensate.
The topological interactions between the magnetic- and electric excitations
in these so-called discrete $H$ gauge theories
are completely described by the Hopf algebra or quantumgroup
$D(H)$.
In 2+1 dimensional space time we may add a Chern-Simons term to
such a model. This deforms the underlying Hopf algebra $D(H)$
into the quasi-Hopf algebra $\DW$ by
 means of a 3-cocycle $\omega$ on $H$.
Consequently, the finite number of physically inequivalent discrete $H$
gauge theories obtained in this way are labelled by the elements of
the cohomology group $H^3(H,U(1))$.  We briefly review
the above results in these notes.
Special attention is given to the Coulomb screening
mechanism operational in the Higgs phase. This mechanism screens the Coulomb
interactions, but not the Aharonov-Bohm interactions.

\end{abstract}
\vspace*{1cm}
Preprint ITFA-93-30, hep-th/9311162.
\newpage  \noindent

\section{Introduction}

By means of the Higgs mechanism
a continuous gauge group  $G$ (for convenience
assumed to be simply connected) of some gauge theory
can be  spontaneously broken down to a
finite residual symmetry group $H$.
It has been known for some time, that
such theories
support magnetic excitations labelled by $\pi_1(G/H)\simeq H$
\cite{niels,bais}.
These magnetic excitations
are stringlike in
3 spatial dimensions and point-like in the
arena in which we will discuss matters,
namely the plain.
Quite recently it has been realized that
besides these magnetic excitations,
the Higgs phase with non-trivial residual symmetry group $H$
also supports charges labelled by the unitary irreducible
representations (UIR's) of $H$~\cite{alf}.
Since the electromagnetic fields are
massive in the Higgs phase, these charges
do not carry  Coulomb fields. They are nevertheless still able
to take part in long range interactions through Aharonov-Bohm
(AB) scattering  with the magnetic excitations~\cite{ver1}.
The physical mechanism behind
this screening of Coulomb charges and
non-screening of AB charges in the Higgs phase was uncovered
in~\cite{morozov}.

The large distance physics of these spontaneously broken
models is described by a
so-called discrete
$H$ gauge theory.
As was shown in Ref.\ \cite{bpm1},
the underlying symmetry algebra is
the Hopf algebra (also called quantumgroup) $D(H)$.
In 2+1 dimensional space time, we
may add a Chern-Simons (CS) term to the action~\cite{des1}.
This deforms the underlying
Hopf algebra $D(H)$ into the quasi-Hopf algebra $\DW$ by
a 3-cocycle $\omega$ on $H$. As a result, there exists a
finite number
of distinct discrete $H$ gauge theories labelled by the
elements of the cohomology group $H^3(H,U(1))$.
These elements are determined by the
CS parameter.

The quasi-Hopf algebra $\DW$  was originally~\cite{dpr1} constructed as
the symmetry algebra of orbifold models~\cite{dvvv} and the related
discrete topological
gauge theories studied in~\cite{diwi}. The  connection between these
models and discrete H gauge theories arising in the Higgs phase, which in some
sense can be viewed as a regularized version of~\cite{diwi}, certainly calls
for further exploration.

Since discrete gauge theories may have emerged after some symmetry breaking
phase transition in the early universe,
our considerations find a context in cosmology~\cite{alf}.
There are also applications in condensed matter systems, such as
nematic crystals~\cite{lizzi}, and type II Landau-Ginzburg superconductors.

 These notes intend to review the results mentioned
above. The outline
is as follows. In section 2 we discuss some
basic features of discrete $H$ gauge theories with CS term.
The example   $H\simeq Z_N$  arising from the symmetry breaking scheme
$SU(2) \rightarrow U(1) \rightarrow Z_N$ will be dealt with in some detail.
We emphasize the difference between the CS screening mechanism
in  unbroken CS electrodynamics and the Higgs screening mechanism
entering the scene when $U(1)$ is spontaneously broken down to $Z_N$.
In both mechanisms Coulomb interactions are screened, while AB
interactions survive.
The peculiarities of non-abelian discrete $H$ gauge theories are briefly
reviewed in section~\ref{ajax1}, whereas the symmetry algebra $\DW$
behind them is discussed in section~\ref{Hopf}.
In section~\ref{zeten} we will apply this machinery to the $Z_N$ gauge
theories. Section~\ref{cat} is dedicated to an
explanation of the notion of Cheshire charges and
Alice fluxes in a $\bar{D}_2$ gauge theory.

\section{Discrete $H$ gauge theories}

As mentioned before, discrete $H$ gauge theories~\cite{alf,bpm1,discr}
naturally arise  whenever
the continuous symmetry group $G$ of some gauge theory is
spontaneously broken down to a finite group $H$ by
the Higgs mechanism.
We will illustrate this scheme starting from a $G \simeq SU(2)$ gauge theory
in (2+1)-dimensional Minkovski space
\be  \label{action}
{\cal L}= -\frac{1}{4}F_{\rho\nu}^a F^{a\,\rho\nu} +
\frac{\mu}{4} \epsilon^{\kappa\sigma\rho}
[F_{\kappa\sigma}^a A^a_\rho + \frac{1}{3}e\epsilon^{abc}A_\kappa^a
A_\sigma^b A_\rho^c ]
+ (D_\rho \Phi)^{\dagger}
\cdot(D^\rho \Phi) - V(\Phi) + {\cal L}_{matter}.
\ee
Greek indices run from 0 to 2, whereas
latin indices label the three (hermitian)
generators of  $SU(2)$.  In our convention the metric $\eta$ has signature
$(+,-,-)$. The covariant
derivative takes the form $D_\rho \Phi =
(\partial_\rho +\imath eA_\rho^a T^a)\Phi$,
with the generators $T^a$ of $SU(2)$  in the representation of the Higgs field
$\Phi$.
In ${\cal L}_{matter}$ we have introduced additional matter fields
minimally coupled to the vector-potential, so that  all
conceivable charge sectors can be discussed.
We have included a CS term in~(\ref{action}) as well~\cite{des1}.
The completely anti-symmetric three dimensional
Levi-Civita tensor
$\epsilon$ appearing in this term is normalized
such that $\epsilon^{012}\equiv 1$.
The demand that the Lagrangian (\ref{action}) should give
rise to a gauge invariant
quantum theory, leads to a quantization condition for the
topological mass  $\mu$~\cite{des1}
\be \label{mu}
\mu = pe^2/4\pi \;\;\;\;\;\; \mbox{with} \;\;\; p\in Z.
\ee

By an appropriate choice of the representation the Higgs field $\Phi$
and the potential $V(\Phi)$,
the gauge symmetry $SU(2)$  can be spontaneously broken  down  to any
finite subgroup $H$ \cite{ovrut}. If we are dealing with energies well below
the symmetry breaking scale,
we are in a Higgs phase with a residual finite gauge symmetry group $H$.
The effective theory we are left with has been called a discrete $H$ gauge
theory~\cite{alf,bpm1,discr}.
It is the purpose of this section to identify the complete spectrum
of charges and magnetic fluxes
of such a theory, together with the topological interactions between them.
We will do so in the simplest example first,
namely the discrete gauge theory that
emerges if we break $SU(2)$ down to $H \simeq Z_N$. This will be the content of
section~\ref{ajax}. The  characteristic features of
discrete gauge theories with a non-abelian residual gauge group $H$
will be briefly reviewed in
section~\ref{ajax1}. Throughout these notes we work with units, such that
$\hbar=c=1$.

\subsection{A $Z_N$ gauge theory} \label{ajax}

One of the interesting features of CS terms is the fact that they endow the
electromagnetic fields with a mass proportional to $\mu$ \cite{des1}.
Thus  charges are screened in the presence of a CS term.
As such, CS terms provide a welcome alternative to
the Higgs mechanism.  The nature of these two screening mechanism is quite
different though. We will contrast the two of them
 in the $U(1)$ phase that arises when the gauge group
$SU(2)$ of~(\ref{action}) is spontaneously broken down to
$U(1)$ at some high energy scale. If the $U(1)$ phase remains unbroken, the
Coulomb screening is due to the CS mechanism, whereas the Higgs mechanism
becomes effective if $U(1)$ is spontaneously broken down to a cyclic
group $Z_N$ at some lower energy scale.

Suppose that the Higgs potential $V(\Phi)$ in (\ref{action})
is
such that the symmetry group $SU(2)$ is spontaneously broken down to $U(1)$
(see for instance \cite{klee} for more details).
The $U(1)$ regime is governed by the following effective Lagrangian
\be                          \label{effaction}
{\cal L}_{ef\!f}= -\frac{1}{4}F_{\rho\nu} F^{\rho\nu} +
\frac{\mu}{4} \epsilon^{\kappa\sigma\rho}F_{\kappa\sigma} A_{\rho}
+({\cal D}_\rho \psi)^*({\cal D}^\rho \psi) - V(|\psi|) + {\cal L}_{matter},
\ee
where $\psi$ denotes an additional Higgs field that we absorbed in ${\cal
L}_{matter}$ in (\ref{action}).
We assume that this Higgs field
$\psi$ carries a global $U(1)$ charge $Ne/2$, i.e.\ ${\cal
D}_{\rho}\psi=(\partial_{\rho}+\im\frac{Ne}{2}A_{\rho})\psi$,
so that we obtain a $Z_N$ gauge theory if this fields condenses at a lower
energy scale~\cite{alf,bpm1,discr}.
The global $U(1)$ charges
are quantized in units of $e/2$ as a consequence of the embedding in $SU(2)$.
In this strictly abelian model, we have omitted the massive modes associated
with the broken generators, and the massive neutral Higgs particles.
Note that this phase also contains instantons labelled by
$\pi_2(SU(2)/U(1))\simeq Z$. In 3 euclidean dimensions
these instantons are monopoles carrying  magnetic
charge $g=4\pi k/e$ with $k\in Z$, while in this (2+1)-dimensional
Minkowski setting they describe quantum tunneling events between states with
magnetic flux difference $|\Delta \phi|=4\pi/e$.
This $U(1)$ gauge theory
is spontaneously broken down to $Z_N$  by endowing the
Higgs field $\psi$ with a non-vanishing vacuum expectation value
$|\!<\! \psi \!>\!|=v$ through the following choice of the
potential
\be                         \label{pot}
V(\mid\psi\mid)=\frac{\lambda}{4}(\mid\psi\mid^2-v^2)^2,  \qquad\qquad
 \lambda, v > 0.
\ee

Before we turn to the subtleties of this spontaneous symmetry
breakdown however, we first consider the unbroken case.
Thus we set $v=0$ for the moment.
Variation  of~(\ref{effaction}) w.r.t.\ to the $U(1)$ vector-potential
$A_{\sigma}$ then yields the following field equation
\be                      \label{fieldequation}
\partial_\rho F^{\rho\sigma} +
\frac{\mu}{2}\epsilon^{\sigma\tau\rho}F_{\tau\rho} = j^\sigma+j^\sigma_H,
\ee
where $j_H^{\sigma} = \im Ne(\psi^*{\cal D}_{\sigma}\psi -
({\cal D}_{\sigma}\psi)^*\psi)$ denotes the Higgs current, and the current
$j^\sigma$  consists of  contributions of the matter fields
contained in ${\cal L}_{matter}$.
Integrating the zeroth component of~(\ref{fieldequation}) over the plain
leads  to  Gauss's law
\be \label{gauss}
Q=q +\mu \phi+q_H=0,
\ee
with $Q=\int\! d^2x\,\vec{\nabla}\!\cdot\!\vec{E}$ the Coulomb charge,
$q=\int\! d^2x \, j^0 $ and $q_H=\int\! d^2x \, j^0_H $
 global $U(1)$ charges, and
$\phi\equiv \int \! d^2x \,\epsilon^{ij}\partial_i A^j$
the total magnetic flux.
The Coulomb charge $Q$ in (\ref{gauss}) vanishes, because the Coulomb fields
carry  a mass $\mu$ in the presence of a CS term, and therefore
vanish exponentially. The screening  mechanism
operating in
unbroken CS electrodynamics attaches fluxes $\phi=-q/\mu$ and
$\phi=-q_H/\mu$ of characteristic size $1/|\mu|$ to the
global $U(1)$ point charges $q$ and $q_H$ respectively~\cite{des1}. This leads
to
an identification of charge and flux at distances
$\gg 1/|\mu|$. The spectrum of this theory at such distances, where
only AB interactions remain between the excitations~\cite{ver1},
is depicted in figure~1.
Note that the AB fields (which are pure gauge)
around the fluxes are still solutions of the field equations. It is not the
vector-potential $A_{\rho}$, that is massive in CS electrodynamics, but rather
the electromagnetic fields~\cite{des1}. The interaction part
$-(j^\rho+j^\rho_H-\frac{\mu}{4}\epsilon^{\rho\kappa\sigma}
F_{\kappa\sigma})A_\rho$
of the Lagrangian~(\ref{effaction}) gives rise to the following
AB phase~\cite{ver1}
\be           \label{ons}
{\cal R}^2 |q_1\!>|q_2\!>=
e^{\im(q_1\phi_2 + q_2\phi_1 + \mu\phi_1\phi_2)}|q_1\!>|q_2\!>=
e^{(-\im\mu\phi_1\phi_2)}|q_1\!>|q_2\!>,
\ee
if we take a charge $q_1$ counterclockwise around $q_2$ once.
This process is effectuated by
the square of the braid operator ${\cal R}$, which interchanges the two
particles in counterclockwise direction.
In the last equality sign,
we used the aforementioned identification of charge and flux.
For the statistical parameter~\cite{leinaas} (generated by ${\cal R}$)
of the excitation with charge $q=-pe$ (with $p \in Z$ defined in (\ref{mu}))
and flux $\phi=4\pi/e$, we
find $\exp\im\theta=\exp(-\im\frac{\mu}{2}\phi^2)=\exp(-2\pi \im p)=1$. So
this excitation is a boson, just as the vacuum, in which it is tunneled
by an instanton.
In fact, this can be seen as an alternative way to derive the quantization
of the CS parameter; in the presence of instantons $\mu$ has to satisfy
(\ref{mu}), otherwise they would tunnel states with different
quantum-statistical properties into each other.
Note that the instantons are charged in the sense that they tunnel between
excitations not only with flux difference $\Delta \phi= -4\pi/e$, but also
with charge difference $\Delta q = pe$ \cite{klee}.
Using~(\ref{ons}),
it is also easily verified that the excitations connected by instantons
are indistinguishable by AB scattering processes with the other excitations
of the spectrum.

\begin{figure}[t]
\begin{center}
\begin{picture}(125,125)(-60,-60)
\put(-60,0){\line(1,0){120}}
\put(0,-60){\line(0,1){120}}
\thinlines
\multiput(-60,0)(10,0){13}{\line(0,1){2}}
\multiput(0,-60)(0,10){13}{\line(1,0){2}}
\multiput(-30,60)(20,-40){4}{\circle{1.6}}
\multiput(-20,40)(20,-40){3}{\circle*{2.0}}
\put(-20,40){\circle{3.6}}
\put(-20,40){\vector(1,-2){20}}
\put(-5,50){\vector(0,1){5}}
\put(43,-3){\vector(1,0){5}}
\put(-12,60){\small$\phi[\frac{\pi}{e}]$}
\put(50,-6){\small$q[\frac{e}{2}]$}
\end{picture}
\vspace{0.5cm}
\caption{\sl The spectrum of CS electrodynamics at distances $\gg 1/|\mu|$.
We depict the flux $\phi$
against the global $U(1)$  charge $q$.
The CS parameter $\mu$ is set to its minimal non-trivial value
$\mu=e^2/4\pi$, i.e.\ $p=1$. The open circles denote half integral charges,
the filled circles integral charges, while
the arrow visualizes the effect of a {\em charged} instanton.}
\end{center}  \label{u1}
\end{figure}
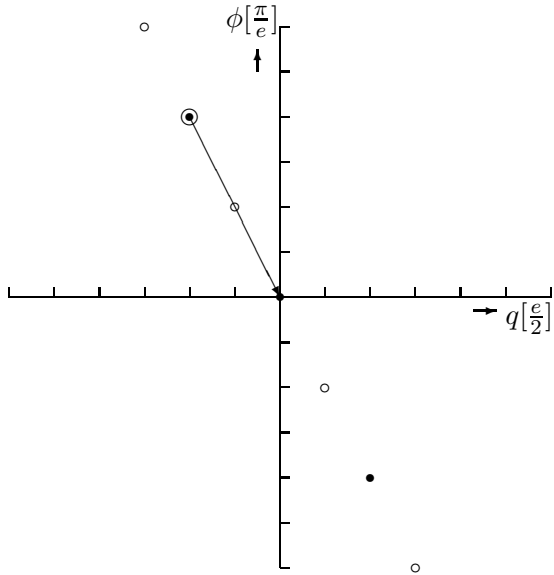

In the broken case ($v \neq 0$) the situation is more involved.
The Higgs field $\psi$ condenses at energy scales lower then
$M_H=v \sqrt{2\lambda}$.
The  $Z_N$ Higgs phase arising at these energy scales
is described by the following simplification in
(\ref{effaction})
\bea  \label{ma}
\mid{\cal D}_{\kappa}\psi\mid^2 & \longrightarrow &
\frac{M_A^2}{2} \tilde A^{\kappa}\tilde A_{\kappa},
\label{mhed}                                       \\
\tilde{A}_{\kappa} & \equiv & A_{\kappa} + \frac{2}{Ne}\partial_{\kappa}
\mbox{Im}\log<\!\psi\!>,
\label{Atilde}
\eea
with $M_A = \frac{Ne}{2}v\sqrt{2}$ and $<\!\psi\!>$ the
vacuum expectation value of $\psi$.
Consequently,
the gauge invariant combination $\tilde{A}$  acquires the
mass~\cite{pisa}
\be
M_{\pm}=\sqrt{M_A^2 + \frac{1}{2}\mu^2 \pm \frac{1}{2}\mu^2\sqrt{
\frac{4 M_A^2}{\mu^2}+1}},
\label{mass}
\ee
where $+$, and $-$ stand for two different components of the photon.

The fact that $\tilde{A}$ is massive does not immediately imply
that $A$ should
also fall off exponentially. It can instead remain pure gauge.
This is  the case
around topological defects of the Higgs condensate
corresponding to  magnetic
vortices~\cite{niels}.
To meet the requirement that the Higgs condensate is
single-valued outside the cores of these vortices,
their  magnetic flux $\phi$  is quantized as
$\phi= 4\pi m/Ne$ with $m \in \pi_1(U(1)/Z_N)\simeq Z$ \cite{niels}.
It can be shown that the magnetic field ($F^1\,_2$) related to the vortices
fall off with $M_-$, and not with $M_+$ ~\cite{boya}.
The shape of this magnetic field depends on the parameters.
For $\mu=0$ we are dealing with
the Abrikosov-Nielsen-Olesen vortex with maximal magnetic field at its centre,
falling off with mass $M_-=M_A$ at large distances~\cite{niels}.
For increasing $|\mu|$ the magnetic field at the centre of the vortex
decreases until it vanishes and becomes a minimum
in the CS limit $e,|\mu| \rightarrow
\infty$, with fixed ratio $e^2/\mu$~\cite{boya}.
(In our case, where the topological mass
$\mu$ is quantized as (\ref{mu}) this limit simply means
$e \rightarrow \infty$ leaving the
CS parameter $p$ fixed). In this CS limit, that boils down to
neglecting the Maxwell term in (\ref{effaction}), the magnetic
field is localized in a ring-shaped region around $1/M_H$~\cite{boya,jacko}.
In the presence of a CS term the  vortices are endowed with further peculiar
properties.
As we see from Eq.\ (\ref{gauss}), a bare flux would imply a non-vanishing
Coulomb charge $Q$, which is inconsistent with the massivity of the
electromagnetic fields. It is the Higgs condensate that brings salvation.
As follows from~(\ref{gauss}) and~(\ref{ma}), at distances $\geq 1/M_H$
it conspires with
the vector-potential to become a charge density $j_{scr}^0$
\be                              \label{jik}
q_H \longrightarrow q_{scr}\equiv \int d^2 x  j_{scr}^0 \equiv
-\int \! d^2 x \, M_A^2 \tilde{A}_0=-\mu\phi,
\ee
establishing the exponential decay of the
Coulomb fields induced by the flux $\phi$ of the
vortex~\cite{morozov}.
This screening charge density $j_{scr}^0$ is localized in a ring outside
the core of the vortex (see \cite{boya} and references given there).
Remarkably enough, the screening charge $q_{scr}$ does not couple to the
AB interaction~\cite{morozov}.
The associated current
$j_{scr}^{\kappa}\equiv -M_A^2\tilde A^{\kappa}$
would only interact with the AB field (produced
by some remote vortex), if there was a term in the Lagrangian of
the form
$-j_{scr}^{\kappa}A_{\kappa}= M_A^2\tilde A^{\kappa}A_{\kappa}$ \cite{ver1}.
Instead we only encounter the  term
$\frac{1}{2} M_A^2\tilde A^{\kappa}\tilde A_{\kappa}$
in  (\ref{ma}). In other words, $q_{scr}$
couples to $\tilde A$ rather than to $A$, and thus
does {\em not} feel AB fields related to remote vortices, which have
non-vanishing $A$-component, but strictly vanishing $\tilde A$ at large
distances from their cores. This implies that  taking
a vortex with flux $\phi_1$ counterclockwise
around another vortex with flux $\phi_2$
(a process denoted by ${\cal R}^2$)  generates
the AB phase $\exp (\im \mu \phi_1 \phi_2)$, entirely due to the coupling
$\frac{\mu}{4}\epsilon^{\rho\kappa\sigma}
F_{\kappa\sigma}A_\rho$ in (\ref{effaction}). Here we assume that the
vortices never overlap.
Identical vortices with flux $\phi$ then behave as
anyons~\cite{leinaas} with
statistical parameter $\exp\im \theta=\exp (\im \mu \phi^2/2)$.
A result which is in complete accordance with the spin-statistics connection
$\exp \im \theta=\exp (2\pi \im s)$, where $s=\mu \phi^2/4\pi$
denotes the spin
that can be calculated for the classical vortex solution~\cite{boya}.

An important issue is
whether vortices will actually form or not, that is
whether the superconductor we are describing here is type II or I
respectively.
In ordinary superconductors ($\mu=0$) an evaluation of the free
energy yields that we are dealing with a type II superconductor if
$M_H/M_A=2\sqrt{\lambda}/Ne \geq 1$, and a type I superconductor otherwise.
A perturbation  analysis for small $\mu \neq 0$ shows that the type
II region is
extended~\cite{jacobs}. In the following, we will always
assume that our parameters are adjusted
such that we are in the type II region.

Let us now turn to the fate of the global $U(1)$ matter charges $q$
in this $Z_N$ Higgs phase. We first consider the situation where
the CS term is absent, that is $\mu=0$. Substituting (\ref{Atilde}) and
(\ref{gauss}) in
(\ref{fieldequation}) yields the  Klein-Gordon equation
for $\tilde{A}^\sigma$, which indicates that the Coulomb fields generated
by the charge  $q$ fall off with mass $M_-=M_A$.  It follows from Gauss's
law (\ref{gauss}) with $\mu=0$,
that this is achieved by surrounding the charge $q$ by the screening
charge density $j_{scr}^0= -M_A^2 \tilde{A}_0$
\be
q_H \longrightarrow q_{scr}\equiv
-\int \! d^2 x \,M_A^2 \tilde{A}_0=-q,
\ee
with support in a ring-shaped region localized at distances $\sim 1/M_H$.
At larger distances, the contribution of
the screening charge $q_{scr}$
to the Coulomb fields  completely
cancels the contribution of $q$.
As we saw before, the screening charge does {\em not} couple
to the AB interaction,
and the AB phase $\exp (\im q\phi)$ generated if $q$ encircles a vortex
$\phi$ in counterclockwise direction won't be canceled by the screening cloud
$q_{scr}$ around $q$~\cite{morozov}. Thus the Coulomb interactions are
exponentially damped
by the Higgs mechanism, while the AB interactions
are not. If we turn on the CS term we have in principle two competing
screening mechanisms. Only the Higgs mechanism can be effective at distances
$\gg 1/M_H$ though. This must be clear already from the fact that
the fluxes $-q/\mu$, attached to the point
charges $q$ by the CS mechanism,
in general do {\em not} satisfy the flux quantization condition
$\phi=4\pi m/Ne$ with $m \in Z$, that arises at these distances.
It is illuminating to illustrate this in the situation where we have
adjusted our parameters such that we are in the CS limit, thus
$e$ is  large with fixed
CS parameter $p$ (therefore the topological mass $|\mu|$ given as (\ref{mu})
is large as well).
As we have argued before, if  $U(1)$ is not spontaneously broken,
i.e.\ $v$ in (\ref{pot}) is sent to $0$, then
the point charges $q$ will be surrounded by fluxes
localized within a region of radius $1/M_+=1/\mu$.
The photon component $M_-=0$ decouples, and
the spectrum at distances $\gg 1/ \mu $  boils down to figure 1.
Now suppose that we turn on the Higgs mechanism
with a small, but non-vanishing value for $v$, i.e.\ $0 < v\ll 1$.
To make sure we are in the type II region, we make the additional assumption
$\lambda\geq
(Ne/2)^2$. So $\mu \gg M_H > M_A$, and in first order
approximation in $(2M_A/\mu)^2$ we find from (\ref{mass}) that
$M_+ \approx \mu$ and  $M_- \approx M_A^2/\mu$. From continuity reasons,
we expect that the charge $q$ will still be surrounded by a flux $-q/\mu$
with support extending to distances $ \sim 1/ \mu$,
so that the Coulomb fields of the point charges $q$ still
fall off with  the $M_+$ component of the photon.
At distances $> 1/M_H$ we have to satisfy the flux quantization condition,
which states that
the total flux must be a multiple of  $4\pi/Ne$.
To achieve this, we have the vortex solution at our disposal.
We expect that in this range of parameters,
the complete solution of the field equations for a point charge $q$
 will be a superposition of that for   unbroken CS electrodynamics
 at short distances $\ll 1/M_H$,
supplemented with a vortex solution with a flux $\phi$ with support
at distances
$1/M_H$,
such that
$-\frac{q}{\mu}+\phi \in 4\pi m /Ne$.
Recall that in the CS limit, the vortex solution is located
in a ring-shaped region with radius  $\sim 1/ M_H$ falling off with
the $M_-$ component at large distances.
Moreover, it becomes trivial at small distances
$\ll 1/ M_H$~\cite{boya,jacko}.
This indicates that we find the spectrum of unbroken
CS electrodynamics (figure~1) at distances $1/\mu \ll r \ll 1/ M_H$,
while the  spectrum of the $Z_N$ Higgs phase (depicted in figure~2 for
$N=4,\, p=1$, and explained in the next paragraphs)
emerges at distances $\gg 1/ M_H$.
It would be interesting to verify this analysis by means of
a numerical evaluation.

The main conclusion from all this is
that in the broken case  the identification
of charge and flux is lost at distances $\gg 1/M_H$;
charge and flux become independent quantum-numbers in the $Z_N$ Higgs phase.
In the $Z_N$ phase the spectrum
 won't reside on a line as in
figure 1, but rather on the lattice spanned by the charge
$q=e/2$ and the flux $\phi=4\pi/Ne$.
We will denote the excitations on this lattice as $|m,n\!>$, where $m$ stands
for
the number of flux units $4\pi/Ne$, and $n$ for the number of charge units
$e/2$. The AB phases generated between these excitations follow
from the coupling $-(j^\rho-\frac{\mu}{4}\epsilon^{\rho\kappa\sigma}
F_{\kappa\sigma})A_\rho$  as
\bea  \label{111}
{\cal R}\;|m_1,n_1\!>\!|m_2,n_2\!> &=&
e^{\im \tilde{Q}_2 \phi_1}|m_2,n_2\!>\!|m_1,n_1\!> \nn \\
 &=& e^{ \im \frac{2\pi}{N} \{ (n_2+p\frac{m_2}{N})m_1)\}}\;
|m_2,n_2\!>\!|m_1,n_1\!>,
\eea
with the charge $\tilde{Q}$ given by
\be    \label{noether}
\tilde{Q} \equiv q + \frac{\mu}{2} \phi.
\ee

There is a large redundancy in the spectrum   as we have sketched it so far.
We have not taken care of  the modulo $N$ calculus yet.
The
proper labelling of the magnetic flux sectors in the full theory is
by $\pi_1(SU(2)/Z_N)\simeq Z_N$ and not by $\pi_1(U(1)/Z_N) \simeq Z$.
The apparent difference can be
understood if the role of the instantons in the model is taken into account.
As mentioned earlier, they connect vortices with a flux difference
$|\Delta \phi| = 4\pi/e$, thus establishing the desired
$Z_N$ calculus. We will argue that this magnetic $Z_N$ calculus
becomes twisted in the presence of a CS term, while the $Z_N$ calculus for the
charges is unaffected.
To that end consider the process in which an
arbitrary composite $(m_3,n_3)$ encircles
a composite $(m_1+m_2,n_1+n_2)$.
The sums  $m_1+m_2$  and $n_1+n_2$
do not necessarily lay between $0$ and $N-1$. Using the notation $[m_1+m_2]$
for $(m_1+m_2)\,\,\,\mbox{modulo}\,\, N $, chosen between $0$ and $N-1$,
we can rewrite the AB phase for this process, generated by ${\cal R}^2$, as
\be \label{modulo}
e^{ \im (\tilde{Q}_3\phi_1 + \tilde{Q}_1 \phi_3 + \tilde{Q}_3\phi_2
+ \tilde{Q}_2 \phi_3 )} = e^{ \im (\tilde{Q}_3\phi_{12} +
\tilde{Q}_{12}\phi_3)},
\ee
with the definitions
\bea
\tilde{Q}_{12} &\equiv& q_{12} + \frac{\mu}{2} \phi_{12} \label{fusQt}\\
   & \equiv &
([n_1+n_2+\frac{2p}{N}(m_1+m_2-[m_1+m_2])] + \frac{p}{N}[m_1+m_2])
\;\frac{e}{2}
 \nn \\
\phi_{12} &\equiv& \frac{4\pi}{Ne}[m_1+m_2].   \label{fusphi}
\eea
The equations~(\ref{fusQt}) and~(\ref{fusphi}) express the way charges
and fluxes `add', i.e.\ they specify the {\em fusion rules}
for the excitations in our model. In terms of the quantumstates
$|m,n\!>$ these read
\be \label{CSfusion}
|m_1,n_1\!>*\,|m_2,n_2\!>=
|[m_1+m_2],[n_1+n_2+\frac{2p}{N}(m_1+m_2-[m_1+m_2])]\!>.
\ee
Hence the spectrum can be confined to an $N$ by $N$ charge/flux lattice.
The modulo $N$ calculus for the fluxes is twisted by the CS parameter $p$
though.
Phrased in more physical terms: the instantons $\Delta \phi=-4\pi/e$ carry
a charge  $\Delta q=pe$, in complete accordance with our findings
 in
unbroken CS electrodynamics.

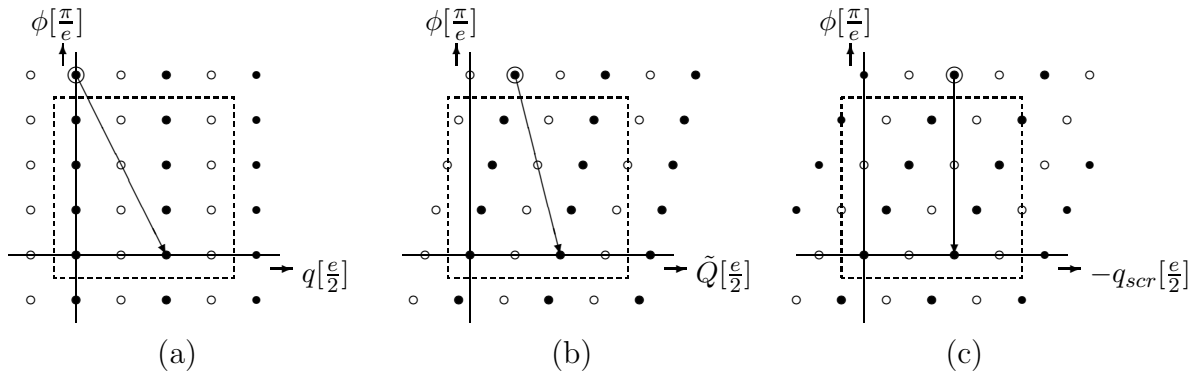
\begin{figure}[t]
\begin{center}
\begin{picture}(85,80)(-15,-15)
\put(-5,-5){\dashbox(40,40)[t]{}}
\put(-15,0){\line(1,0){60}}
\put(0,-15){\line(0,1){60}}
\thinlines
\multiput(-10,-10)(0,10){6}{\multiput(0,0)(10,0){5}{\circle{1.6}}}
\multiput(0,-10)(0,10){6}{\multiput(0,0)(20,0){3}{\circle*{2.0}}}
\put(0,40){\circle{3.6}}
\put(0,40){\vector(1,-2){20}}
\put(-3,42){\vector(0,1){5}}
\put(43,-3){\vector(1,0){5}}
\put(-10,50){\small$\phi[\frac{\pi}{e}]$}
\put(50,-6){\small$q[\frac{e}{2}]$}
\put(18,-24){(a)}
\end{picture}
\begin{picture}(85,80)(-15,-15)
\put(-5,-5){\dashbox(40,40)[t]{}}
\put(-15,0){\line(1,0){60}}
\put(0,-15){\line(0,1){60}}
\thinlines
\multiput(-12.5,-10)(2.5,10){6}{\multiput(0,0)(10,0){6}{\circle{1.6}}}
\multiput(-2.5,-10)(2.5,10){6}{\multiput(0,0)(20,0){3}{\circle*{2.0}}}
\put(-3,42){\vector(0,1){5}}
\put(43,-3){\vector(1,0){5}}
\put(10,40){\circle{3.6}}
\put(10,40){\vector(1,-4){10}}
\put(-10,50){\small$\phi[\frac{\pi}{e}]$}
\put(50,-6){\small$\tilde{Q}[\frac{e}{2}]$}
\put(18,-24){(b)}
\end{picture}
\begin{picture}(85,80)(-15,-15)
\put(-5,-5){\dashbox(40,40)[t]{}}
\put(-15,0){\line(1,0){60}}
\put(0,-15){\line(0,1){60}}
\thinlines
\multiput(-15,-10)(5,10){6}{\multiput(0,0)(10,0){5}{\circle{1.6}}}
\multiput(-5,-10)(5,10){6}{\multiput(0,0)(20,0){2}{\circle*{2.0}}}
\multiput(-15,10)(5,10){4}{\multiput(0,0)(20,0){1}{\circle*{2.0}}}
\multiput(35,-10)(5,10){4}{\multiput(0,0)(20,0){1}{\circle*{2.0}}}
\put(-3,42){\vector(0,1){5}}
\put(43,-3){\vector(1,0){5}}
\put(20,40){\circle{3.6}}
\put(20,40){\vector(0,-1){40}}
\put(-10,50){\small$\phi[\frac{\pi}{e}]$}
\put(50,-6){\small$-q_{scr}[\frac{e}{2}]$}
\put(18,-24){(c)}
\end{picture}
\vspace{0.5cm}
\caption{\sl The spectrum of a Higgs phase with residual gauge group $Z_4$.
We depict the flux $\phi$
against the global $U(1)$ charge $q$, the  Noether charge $\tilde{Q}$
and the screening charge  $-q_{scr}=q+\mu\phi$ respectively. The
CS parameter $\mu$ is set to its minimal non-trivial value
$\mu=e^2/4\pi$, i.e.\ $p=1$. The identification of the encircled excitation
with an excitation inside the dashed box is indicated with an arrow.
As in figure 1
for unbroken CS electrodynamics, this
arrow visualizes the effect of a {\em charged} instanton. }
\end{center}
\end{figure}

It is clarifying to summarize the foregoing discussion
in pictures, as is done in figure 2 for a $Z_4$ gauge theory.
{}From these pictures
it is immediate that for odd $N$ the fusion algebra $Z_N \times Z_N$
of a discrete $Z_N$ gauge theory without CS term, is altered into
$Z_{kN} \times Z_{N/k}$ in the presence of a CS term~\cite{dvvv}.
Here we defined $k \equiv N/(p,N)$ with $(p,N)$ the greatest common
divisor of the CS parameter $p$ and $N$. So in particular for $p=1$,
the complete spectrum is generated by a single excitation.
For even $N$ we find a similar result, except that the formula
for $k$ has to be replaced by $k \equiv N/(2p,N)$.

The symmetry algebra behind this spectrum is the Hopf algebra
$D^{\omega}(Z_N)$~\cite{bpm1}.
It consists of $Z_N$ gauge transformations and projection operators
signalling the flux of a state. We will denote the elements of this
algebra by $\hook{m}{l}$ ($m,l \in 0,\ldots,N-1$), which
performs a gauge transformation with parameter $l$, and subsequently projects
the state on the flux $m$
\be \label{ac+co}
\hook{m}{l}|m_1,n_1\!>\,=
\delta_{m,m_1} e^{ \im \frac{2\pi}{N}l(n_1+p\frac{m_1}{N})}\;|m_1,n_1\!>.
\ee
Note that it is the charge $\tilde{Q}$~(\ref{noether}), which appeared in
the braid process~(\ref{111}), that gives the
response of a state to a gauge transformation~\cite{bpm1}. A braid process
has the same effect as a gauge transformation.
We leave a more detailed description of the symmetry algebra for
section~\ref{Hopf}.

\subsection{Non-abelian discrete gauge theories} \label{ajax1}

Of course we may also be left with a non-abelian finite symmetry
group $H$ of the Higgs condensate~\cite{ovrut}.
The discussion of such non-abelian discrete gauge theories is slightly more
involved.
As  in the abelian case,
there are three types of excitations in these theories, namely
purely magnetic vortices, AB charges and of course
the dyons that are composites of these last two excitations.
If we assume that $G$ is simply connected,
the magnetic vortices are labelled~\cite{bais} by
the elements of $\pi_1(G/H) \simeq H$. They are stringlike in
3 spatial dimensions, while they are particle-like in the plain.
A residual gauge transformation $g \in H$ acts on the magnetic
charges $h \in H$ through
 conjugation
\be                                     \label{haro}
g: \qquad |h\!> \longrightarrow |ghg^{-1} \!>.
\ee
So the gauge invariant labelling of the magnetic
charges is by means of the conjugacy classes $^A C$ of $H$.
We have to keep in mind however, that physical properties
such as braiding~\cite{bais,alf,bpm1,discr}
\be             \label{metamo}
\qquad {\cal R}\; |h \!>|k \!>\; = |hkh^{-1}\!>|h\!>,
\ee
depend on the specific element of $^A C$ by
which the magnetic flux is represented. As before, the effect of
braiding and gauge transformations is similar.

The free electric charges are labelled by
the UIR's of the gauge group $H$~\cite{alf}. In the presence of a
magnetic flux $h \in\,\! ^A C$ the gauge group $H$ is broken down
to the centralizer $^h N$ of $h$ in $H$ though~\cite{alf,lizzi}, and
the electric charges we can put on the magnetic flux $h$ are labelled
by the UIR's of $^h N$~\cite{bpm1}.
If we now use the fact that the centralizers
of the different $h \in \,\! ^A C$ are isomorphic, then  we can summarize
the superselection sectors of a discrete $H$ gauge theory as
\[
(^A C, ^\alpha \Gamma),
\]
with $^\alpha \Gamma$ an UIR of the centralizer $^AN$ associated
with the conjugacy class $^AC$. As we shall see in the next section,
these superselection
sectors exactly coincide with the irreducible representations of
the Hopf algebra  $D(H)$, and it turns out
that the topological properties of the
excitations in discrete $H$ theories can be completely described in terms of
this algebra, and its representation theory~\cite{bpm1}.
If we add a CS term to the action
of these theories, then  the underlying Hopf algebra $D(H)$  is deformed
into the quasi-Hopf algebra $\DW$ by means of a 3-cocycle $\omega$  on $H$.

\section{The quasi-Hopf algebra $\DW$}   \label{Hopf}

The quasi-triangular quasi-Hopf algebra $\DW$ was first discussed  by
Dijkgraaf, Pasquier and Roche~\cite{dpr1} and we will adopt their notation.
For a general introduction
into the notion of quasi-triangular quasi-Hopf algebras we refer to the work of
Drinfeld~\cite{dri1}.

The symmetry algebra $\DW$ behind discrete $H$ gauge theories
is spanned by the elements $\{\hook{g}{x}\}_{g,x\in H}$,
denoting a residual gauge transformation $ x \in H$ followed
by a projection on the flux state $|g\!>$.  In terms
of these basis elements the multiplication, the co-multiplication $\Delta$,
the associator $\varphi$ and the $R$-matrix
read
\bea
\hook{g}{x}.\hook{h}{y} &=& \delta_{g,xhx^{-1}}\hook{g}{xy} \theta_g(x,y)
\label{algebra}\\
 \Delta(\,\hook{g}{x} \,) &=& \sum_{\{h,k|hk=g\}}\hook{h}{x}\otimes\hook{k}{x}
\gamma_x(h,k) \label{coalgebra}
\\
\varphi &=& \sum_{g,h,k}\,\omega^{-1}(g,h,k)\,
\hook{g}{e}\otimes\hook{h}{e}\otimes\hook{k}{e}  \label{isom} \\
R &=&\sum_{g,h}\,\hook{g}{e}\otimes\hook{h}{g} \; ,  \label{Rmatrix}
\eea
where $\theta,\gamma$ and $\omega$ are phases that
equal 1 whenever one of their variables is the unit $e$ of $H$. The algebra
morphism $\Delta$ from $\DW$ to $\DW \ot \DW$ enables us to construct the
tensorproduct of representations $(\Pi_1, V_1)$ and $(\Pi_2, V_2)$ of $\DW$,
and therefore  to extend the action of the symmetry algebra from
1-particle states to  2-particle states.
The associator $\varphi$ establishes the isomorphism
$\Pi_1 \ot \Pi_2 \ot \Pi_3(\varphi)$ between the
representation-spaces $(V_1 \ot V_2) \ot V_3$  and
$V_1 \ot (V_2 \ot V_3)$, constructed by $(\Delta \ot id)\Delta$ and
$(id \ot \Delta)\Delta$ respectively.
The $R$-matrix describes the braiding properties of the particles,
as will become clear later on.

For a consistent implementation~\cite{dpr1,dri1}
of $\varphi$ on tensorproducts of
four representations $\omega$ has to satisfy the 3-cocycle
condition
\be
\label{pentagon}
\delta\omega(g,h,k,l)=\frac{\omega(g,h,k)\;\omega(g,hk,l)\;\omega(h,k,l)}
{\omega(gh,k,l)\;\omega(g,h,kl)}=1,
\ee
where $\delta$ denotes the coboundary operator (interested readers
are referred to \cite{bpm1} for more details on the cohomology structure
appearing here).
So $\omega$ is an element of the cohomology group
$H^3(H,U(1)$. This element is determined by the CS parameter $p$~\cite{bpm1}.
To proceed, the phases $\theta$ and $\gamma$ are completely
prescribed by $\omega$~\cite{dpr1}. This implies that
$\theta$ is a conjugated 2-cocycle
\be
\label{theta}
\tilde{\delta}\theta_{g}(x,y,z)=\frac{\theta_{x^{-1}gx}(y,z)\,\theta_g(x,yz)}
{\theta_{g}(x,y) \,\theta_{g}(xy,z)} = 1,
\ee
with $\tilde{\delta}$ the `conjugated' coboundary operator.
Eq.\ (\ref{theta}) expresses associativity of the
multiplication~(\ref{algebra}). It is also easily verified that
the co-multiplication is quasi-coassociative.

The irreducible representations of $D^{\omega}\!(H)$,
which label the excitations of a
discrete $H$ gauge theory, can be found
by inducing the unitary irreducible representations (UIR's) of the centralizer
subgroups.
Let $\{\,^A\!C\}$ be the set of conjugacy classes of $H$
and introduce a fixed but arbitrary ordering $^A\!C= \{^A\!g_1,\;^A\!g_2,
\ldots,\,^A\!g_k\}$. Let $^A\!N$ be the centralizer of $^A\!g_1$ and
$\{^A\!x_1,\,^A\!x_2,\ldots,\,^A\!x_k\}$ be a set of representatives of the
equivalence classes of $H/^A\!N$, such that $^A\!g_i=\,^A\!x_i\,^A\!g_1\,
^A\!x_i^{-1}$. Choose for convenience $^A\!x_1=e$.
Now consider the
complex vectorspace $V^A_{\alpha}$ spanned by the basis
$\{|^A\!g_j,\,^{\alpha}\!v_i\!>\}_{j=1,\ldots,k}^{i=1,\ldots,
\mbox{\scriptsize dim}^{\alpha}\!\Gamma}$,  where $^{\alpha}\!v_i$ denotes
a basis element of the
UIR  $^{\alpha}\!\Gamma$  of $^A\!N$.
This vectorspace carries an irreducible representation
$\Pi^A_{\alpha}$ of $\DW$ given by
\be \label{13}
\Pi^A_{\alpha}(\,\hook{g}{x})|\,^A\!g_i,\,^{\alpha}\!v_j \!>=
\delta_{g,x\,^A\!g_ix^{-1}}\;\; \varepsilon_g(x)\;|x\,^A\!g_ix^{-1},
\,^{\alpha}\!\Gamma(\,^A\!x_k^{-1}x\,^A\!x_i)\,^{\alpha}\!v_j \!>,
\ee
with $\,^A\!x_k$  defined through $\,^A\!g_k \equiv x\,^A\!g_ix^{-1}$. The new
ingredient here is the phase $\varepsilon_g$ that is
related to $\theta_g$ by
\be \label{repphase}
\theta_g(x,y)=\tilde{\delta} \varepsilon_g (x,y)=
\frac{\varepsilon_g(x)\varepsilon_{x^{-1}gx}(y)}{\varepsilon_g(xy)} \;\; ,
\ee
in order to make~(\ref{13}) a representation.
Note  that~(\ref{repphase}) is equivalent to
the statement that $\theta_g$ is a `conjugated' 2-coboundary,
a property that is not automatically
assured by~(\ref{theta}).
In passing, we also mention that it can be shown (e.g.\ \cite{bpm1})
that the representation theory of $\DW$ indeed only depends on the
cohomology class of $\omega$ and not on the representative we choose in
such a class.

We turn to the fusion rules of $\DW$. Let $\Pi^A_{\alpha}$ and
$\Pi^B_{\beta}$ once again denote irreducible representations of $\DW$.
The tensorproduct representation $\Pi^A_{\alpha} \ot \Pi^B_{\beta}$ defined
by means of the co-multiplication~(\ref{coalgebra}),
need not be irreducible. In general, it  gives rise to a
decomposition
\be               \label{piet}
\Pi^A_{\al}\otimes\Pi^B_{\beta}=\sum_{C\gamma}
N^{AB\gamma}_{\alpha\beta C} \Pi^C_{\gamma},
\ee
with $N^{AB\gamma}_{\alpha\beta C}$  the multiplicity of the irreducible
representations $\Pi^C_{\gamma}$. In other words, the tensorproduct
representation $\Pi^A_{\alpha} \ot \Pi^B_{\beta}$ is completely reducible.
Relation (\ref{piet}) is called a
fusion rule of  $\DW$. Phrased physically: it determines which excitations
$(C,\gamma)$ can be formed in {\em non}-elastic scattering
processes among an excitation
$(A,\alpha)$ and an excitation $(B,\beta)$.
The fusion algebra spanned by the elements
$\Pi^A_{\alpha}$ with multiplication rule~(\ref{piet}), is
commutative and associative. It can therefore be diagonalized by a single
matrix, the so-called modular $S$ matrix.
For $\DW$, this matrix takes the form~\cite{dvvv}
\be                                \label{fusion}
S^{AB}_{\alpha\beta}=\frac{1}{|H|}\sum_{\stackrel{\,^A\!g_i\in\,^A\!C\,,^B\!g_j\in\,
^B\!C}{[\,^A\!g_i,\,^B\!g_j]=e}}
\alpha^*(\,^A\!x_i^{-1}\,^B\!g_j\,^A\!x_i)\beta^*(\,^B\!x_j^{-1}\,^A\!g_i\,
^B\!x_j)\,\,\sigma(\,^A\!g_i | \,^B\!g_j),
\ee
with $\alpha(g)\equiv \mbox{tr}\;^{\alpha}\Gamma(g)$ and the phase
$\sigma(g|h) \equiv \varepsilon_g(h)\varepsilon_h(g)$. We denote the order of
the group $H$ by $|H|$.
The modular $S$ matrix contains all information about the fusion algebra.
In particular, the multiplicities $N^{AB\gamma}_{\alpha\beta C}$
can be obtained
from the modular $S$ matrix by means of Verlinde's formula~\cite{ver0}
\be      \label{verlinde}
N^{AB\gamma}_{\alpha\beta C}=\sum_{D,\delta}\frac{
S^{AD}_{\alpha\delta}S^{BD}_{\beta\delta}
(S^{*})^{CD}_{\gamma\delta}}{S^{eD}_{0\delta}}.
\ee

The elastic scattering processes between the excitations are
governed by the braid operator ${\cal R}$, associated
with the $R$ matrix~(\ref{Rmatrix})
\be \label{whynot}
{\cal R}_{\alpha\beta}^{AB}\equiv
\sigma\circ(\Pi_{\alpha}^A\otimes\Pi_{\beta}^B)(R),
\ee
where $\sigma$ effectuates a permutation.
To be explicit, the braid operation ${\cal R}$ on the state
$|\,^A\!g_i,\,^{\alpha}\!v_j\!>|\,^B\!g_k,\,^{\beta}\!v_l\!>
\in
V^A_{\alpha}\ot V^B_{\beta}$
reads
\be                 \label{braidaction}
{\cal R}_{\alpha\beta}^{AB}\;|^A\!g_i,\,^{\alpha}\!v_j\!>\!
|^B\!g_k,\,^{\beta}\!v_l\!> =
\! \varepsilon_{^B\!g_m}(\,^A\!g_i)
|\,^B\!g_m,
\,^{\beta}\!\Gamma(^B\!x_m^{-1\;A}\!g_i^B\!x_k)^{\beta}\!v_l\!> |^A\!g_i,
^{\alpha}\!v_j\!>,
\ee
where $^B\!x_m$ is defined through $^B\!g_m \equiv
\,^A\!g_i\,^B\!g_k\,^A\!g_i^{-1}$. Note that~(\ref{braidaction})
incorporates the braid properties found in~(\ref{111}) and~(\ref{metamo}).
For a non-trivial 3-cocycle $\omega$, the braid operator ${\cal R}$
does {\em not} satisfy the ordinary -,
but rather the quasi Yang-Baxter equation
\be \label{quasiYBE}
{\cal R}_1  \tilde{\cal R}_2 {\cal R}_1=
\tilde{\cal R}_2   {\cal R}_1  \tilde{\cal R}_2  \; .
\ee
 ${\cal R}_1$ acts on the three-particle states in the space
$(V_1 \ot V_2) \ot V_3$
as ${\cal R}\otimes {\bf 1}$ and $\tilde{\cal R}_2$ as
$\Phi^{-1}\cdot({\bf 1}\otimes {\cal R})\cdot\Phi$ with
$\Phi\equiv \Pi_1 \ot \Pi_2 \ot \Pi_3 (\varphi)$,
$\Phi^{-1}\equiv \Pi_2 \ot \Pi_1 \ot \Pi_3 (\varphi^{-1})$ and $\varphi$ the
associator~(\ref{isom}).

The cross sections of elastic two-particle Aharonov-Bohm scattering
are completely determined by the monodromy matrix ${\cal R}^2$
\be \label{Aharonov}
\frac{d\sigma}{d\varphi}=\frac{1}{2\pi k
\sin^2(\varphi/2)}\;\frac{1}{2}\,[1-\mbox{Re}<\!\psi_{in}|{\cal
R}^2|\psi_{in}\!>],
\ee
with $|\psi_{in}\!>$  the incoming two-particle state, and $k$
the relative momentum (recall that we are working with natural units
$\hbar=c=1$)~\cite{ver1,bpm1,preslo}.

\subsection{$Z_N$ gauge theories revisited}      \label{zeten}

We briefly illustrate these rather mathematical considerations
by approaching the $Z_N$ gauge theories (section~\ref{ajax}) from the
Hopf algebra point of view.

Every element $m \in Z_N$ constitutes a conjugacy
class and has the full group $Z_{N}$ as its centralizer.
The $N$ different UIR's $^l \Gamma$ of $Z_N$
are all 1-dimensional and  are given by
$^l \Gamma(m) = e^{ \im \frac{2\pi}{N}lm}$.
Thus the  representations of $D^{\omega}(Z_N)$ in turn can be labelled as
$\Pi_n^m \equiv (m,n)$, where  $m$ denotes an element
of $Z_N$ and $n$ the $Z_N$ representation
$^n  \Gamma$. With $m$ the number of flux units $4\pi/Ne$, and
$n$ the charge in units $e/2$, this is exactly the spectrum we
found for a $Z_N$ gauge theory.

It is well-known~\cite{cartan}  that all
$H^{even}(Z_{N},U(1))\simeq 1$, while $H^{odd}(Z_{N},U(1))\simeq Z_{N}$.
An  explicit realization of
$\omega\in H^3(Z_N,U(1))$ is given by
\be \label{omom}
\omega(m_1,m_2,m_3)= e^{ \im \frac{2\pi p}{N^2}m_1(m_2+m_3-[m_2+m_3])},
\ee
with $p \in [0,\ldots,N-1]$.
It is easily inferred~\cite{bpm1} that  $\theta_{m_1}(m_2,
m_3)=\gamma_{m_1}(m_2,m_3)=\omega(m_1,m_2,m_3)$.
Consequently relation~(\ref{repphase}) is solved by
\be\label{ep1}
\varepsilon_{m_1}(m_2)=e^{\im \frac{2\pi p}{N^2}m_1m_2}.
\ee
With the help of~(\ref{ep1}) we now obtain~(\ref{ac+co}) from~(\ref{13}),
(\ref{CSfusion}) from~(\ref{verlinde}) and~(\ref{111})
from~(\ref{braidaction}).

Note that in this abelian example we find
$\tilde{\cal R}_2={\cal R}_2$, because of the symmetry of $\omega$ in the last
two entries:
$\omega(m_1,m_2,m_3)=\omega(m_1,m_3,m_2)$.
This implies
that the  quasi Yang-Baxter equation~(\ref{quasiYBE}) projects down
to the ordinary Yang-Baxter equation ${\cal R}_1 {\cal R}_2 {\cal R}_1=
{\cal R}_2 {\cal R}_1{\cal R}_2$. To proceed, the
Aharonov-Bohm cross-sections~(\ref{Aharonov}) takes the form
\be \label{crosssections}
\frac{d\sigma}{d\varphi}[(m_1,n_1),(m_2,n_2)]=
\frac{\sin^2 \frac{\pi}{N}(n_1 m_2+ n_2 m_1 +\frac{2p}{N}m_1 m_2)}
{2\pi k \sin^2(\varphi/2)}\;,
\ee
where the factor $\frac{1}{\sin^2 \varphi/2}$ has to be replaced by
$\frac{1}{\sin^2 \varphi/2} + \frac{1}{\cos^2 \varphi/2}$ if the
two particles are indistinguishable~\cite{bpm1}.

\subsection{Cheshire charges and Alice fluxes}  \label{cat}

In a non-abelian setting intriguing phenomena as Alice fluxes
and Cheshire charges arises~\cite{alf,bpm1,discr}.
We briefly discuss these phenomena
in a $\bar{D}_2$ gauge theory.

The dihedral group $\bar{D}_2$ is of order $8$ with 5 conjugacy classes
$\{ e \}$, $\{ \bar{e} \}$, $\{X_1,\bar{X_1}\}$, $\{X_2,\bar{X_2}\}$,
$\{X_3,\bar{X_3}\}$. There are four 1 dimensional UIR's of $\bar{D}_2$,
and one 2-dimensional UIR. We will label the trivial UIR  by $1$, while
the other 1-dimensional UIR's will be denoted by $J_a$, $a=1,2,3$. The UIR
$J_a$ sends all elements into $-1$ except for $e,\bar{e},X_a,\bar{X}_a$,
which are represented by $1$.
The 2-dimensional UIR labelled by $\phi$ is obtained by
sending $X_a/\bar{X}_a$ into $\pm \im \sigma_a$ with $\sigma_a$ the Pauli
matrices. This is the purely electric part of the spectrum of our
$\bar{D}_2$ gauge theory. Note that $\bar{e}$ also got the full group
$\bar{D}_2$ as its centralizer. We label these dyonic excitations by
overlining the UIR's of $\bar{D}_2$. The $X_a$ conjugacy classes have
a $Z_4$ centralizer. In the remainder, we will only work with
the purely magnetic fluxes, labelled by $\sigma_a^+$.

For the cocycle structure we use the general result
 $H^3(H,U(1)\simeq Z_{|H|}$ for
subgroups $H$ of $SU(2)$ \cite{cartan}. Recall that
$|H|$ denotes the order of $H$ . In our case this leads us to
$H^3(\bar{D}_2)\simeq
Z_8$, so there are in principle 8 different $\bar{D}_2$ gauge theories.
A numerical solution of the cocycle structure shows
that there are only 4 different sets of fusionrules though,
i.e.\ in terms of the fusionrules the CS parameter $p$ is periodic
with period 4 (see table \ref{fuzzy} and Ref.\ \cite{bpm1}).

\begin{table}[h]
\begin{center}
\begin{tabular}{cccccc}  \hline
fusion rule           &    &    p=0      &   p=1    &     p=2   &     p=3    \\
\hline
$J_a * J_a$           & &  $1$ & $1$ & $1$ & $1$  \\
$\phi * \phi$         & &  $1+\sum_b J_b$ & $1+\sum_b J_b$ & $1+\sum_b J_b$ &
 $1+\sum_b J_b$ \\
$\sigma^{+}_a*\sigma^{+}_a$ & &$1+J_a+\bar{1}+\bar{J}_a$&
$1+J_a+\bar{\phi}$& $1+J_a+\sum_{c\neq a}\bar{J}_c$&$1+J_a+\bar{\phi}$ \\
             \hline
\end{tabular}
\end{center}
\caption{\sl The fusion rules
(for different values of the CS parameter $p$) that play a role in the process
depicted in figure 3. The fusion rules are
periodic in $p$ with period $4$.}
\label{fuzzy}
\end{table}

Now let us consider the process in which we start from the vacuum $1$ and
at some time create a flux/anti-flux pair $\sigma_a^+$
and a charge/anti-charge pair $\phi$.
Note that this is possible for all values of the CS parameter $p$,
since the vacuum appears in the fusionrules $\phi * \phi$ and
$\sigma_a^+ * \sigma_a^+$ (displayed in table~\ref{fuzzy}) irrespective of
the value of $p$.
Subsequently, the charge
$\phi$ is taken around the flux $\sigma^+_a$, and fused with the other member
of the $\phi$ pair again.
In terms of quantumstates, this process (depicted in figure 3) reads
\bea   \label{uit}
1 & \longrightarrow &   1 \ot 1    \\
 & \longrightarrow & \frac{1}{2}
[ |X_1\!>|\bar{X}_1\!> +|\bar{X}_1\!>|X_1\!>] \ot
[|^{\phi}v_1\!>|^{\phi}v_2\!> -|^{\phi}v_2\!>|^{\phi}v_1\!>]  \nn \\
  & \stackrel{{\bf 1} \ot {\cal R}^2 \ot {\bf 1}}{\longrightarrow} &
 \frac{1}{2}\{ \im [|\bar{X}_1\!>|X_1\!> - |X_1\!>|\bar{X}_1\!>]\ot
  [|^{\phi}v_2\!>|^{\phi}v_2\!> -|^{\phi}v_1\!>|^{\phi}v_1\!>] \} \nn \\
   & \longrightarrow & J_1 \ot J_1  \nn \\
  & \longrightarrow & 1,            \nn
\eea
where $^{\phi}v_1=(1,0)$ and $^{\phi}v_2=(0,1)$ are the basisvectors
of the 2 dimensional UIR $\phi$, in which the elements $X_a/\bar{X}_a$
of $\bar{D}_2$ are
represented by  $\pm \im \sigma_a$ with $\sigma_a$ the Pauli
matrices. For convenience we only consider $a=1$. After the $\phi$
charge has encircled the flux, the nature of the charge pair has changed.
It is easily verified that the quantumstate describing the $\phi$ pair
behaves as a $J_1$ charge under $\bar{D}_2$ gauge transformations,
instead of the trivial charge~$1$.
This is the reason why $\sigma_1^+$ is also called an Alice flux~\cite{alf}.
Going around an Alice flux with a charge has a similar effect
as going through the looking glass in Lewis Carroll's  `Alice's Adventures in
Wonderland'. Actually, the analogy with these adventures can be pushed
further. The charge $J_1$ of the $\phi$  pair is not an ordinary
charge. It's a property of the pair, it can {\em not} be localized
on the constituents of the pair nor anywhere else,
just as the elusive smile of the Cheshire cat in `Alice's Adventures
in Wonderland'.
Therefore the name Cheshire charge has been coined~\cite{alf}.
After amalgamating the members of the $\phi$

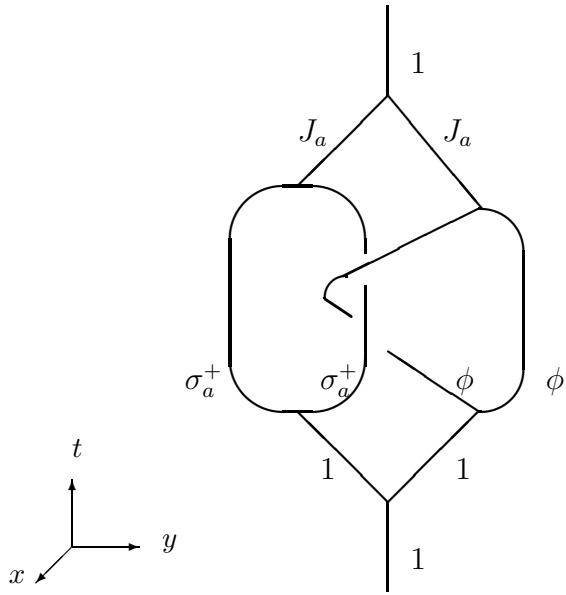
\begin{figure}[bh]
\begin{center}
\begin{picture}(125,140)(-60,-60)
\thicklines
\put(0,-60){\line(0,1){20}}
\put(0,-40){\line(1,1){20}}
\put(0,-40){\line(-1,1){20}}
\put(-20,-5){\oval(30,30)[b]}
\put(-20,15){\oval(30,30)[t]}
\put(-35,-10){\line(0,1){25}}
\put(-5,-10){\line(0,1){18}}
\put(20,-5){\oval(20,30)[rb]}
\put(20,10){\oval(20,30)[rt]}
\put(20,25){\line(-2,-1){30}}
\put(20,-20){\line(-3,2){20}}
\put(-9,5){\oval(10,10)[lt]}
\put(-14,5){\line(3,-2){6}}
\put(30,-10){\line(0,1){20}}
\put(0,50){\line(-1,-1){20}}
\put(0,50){\line(5,-6){21}}
\put(0,50){\line(0,1){20}}
\put(5,-55){$1$}
\put(15,-35){$1$}
\put(-15,-35){$1$}
\put(-45,-15){$\sigma_a^+$}
\put(35,-15){$\phi$}
\put(15,-15){$\phi$}
\put(-15,-15){$\sigma_a^+$}
\put(5,55){$1$}
\put(12,40){$J_a$}
\put(-20,40){$J_a$}
\thinlines
\put(-70,-50){\vector(1,0){15}}
\put(-50,-50){\small $y$}
\put(-70,-50){\vector(0,1){15}}
\put(-70,-30){\small $t$}
\put(-70,-50){\vector(-1,-1){8}}
\put(-84,-58){\small $x$}
\end{picture}
\vspace{0.5cm}
\caption{\sl
After the $\phi$ charge has encircled the Alice flux $\sigma_a^+$,
the flux/anti-flux pair $\sigma_a^+$ and charge/anti-charge $\phi$
carry Cheshire charge $J_a$.}
\end{center}  \label{cheshire}
\end{figure}
\newpage
\noindent pair the Cheshire charge
$J_1$ becomes an ordinary localized $J_1$ charge.
Similar observations appear for the flux pair. If we recall (see~(\ref{haro}))
that the
gauge group $\bar{D}_2$ acts by means of conjugation on the fluxes,
we see that the $\sigma_1^+$ pair is also endowed with a Cheshire charge
$J_1$ after the  charge $\phi$ has encircled the Alice flux
$\sigma_1^+$.  This Cheshire charge becomes a localized $J_1$ charge upon
fusing the members of the flux pair, and as table~\ref{fuzzy} indicates
the two $J_1$ charges we are left with now,
can be annihilated  into the vacuum.
So this local process did not alter any global properties, just as it should.
Note that the cocycles do {\em not} enter this braid process involving the pure
charge $\phi$. It only enters braid processes among fluxes, as is clear
from the $Z_N$ example.

\subsection*{Acknowledgments}
M. de W.P. would like to thank the organizers for their kind invitation,
and for the enormous hospitality he has
enjoyed in Moscow and Alushta. He is especially grateful to the people
that came for his rescue in the infamous restaurant incident.

\end{document}